\documentclass[10pt]{iopart}
\usepackage{graphicx}
\usepackage{cite}
\pdfoutput=1

\begin{document}

\title[]{Versatile transporter apparatus for experiments with optically trapped Bose-Einstein condensates}

\author{Daniel~Pertot, Daniel~Greif\footnote{Present
address: Institute for Quantum Electronics, ETH Z{\"u}rich, 8093 Z{\"u}rich,
Switzerland.}, Stephan~Albert\footnote{Present address: Physik Department,
Technische Universit{\"a}t M{\"u}nchen, 85748~Garching, Germany.},
Bryce~Gadway, and Dominik~Schneble}
\address{Department of Physics and Astronomy, Stony Brook University,\\ Stony~Brook, NY~11794, U.S.A.}

\ead{dpertot@ic.sunysb.edu}

\begin{abstract}
We describe a versatile and simple scheme for producing magnetically and
optically-trapped $^{87}$Rb Bose-Einstein condensates, based on a moving-coil
transporter apparatus. The apparatus features a TOP trap that incorporates the
movable quadrupole coils used for magneto-optical trapping and long-distance
magnetic transport of atomic clouds. As a stand-alone device, this trap allows
for the stable production of condensates containing up to one million atoms. In
combination with an optical dipole trap, the TOP trap acts as a funnel for
efficient loading, after which the quadrupole coils can be retracted, thereby
maximizing optical access. The robustness of this scheme is illustrated by
realizing the superfluid-to-Mott insulator transition in a three-dimensional
optical lattice.
\end{abstract}

\pacs{67.85.-d, 37.10.Gh, 67.85.Hj}

\maketitle

\section{Introduction}

Techniques for the optical trapping and manipulation of Bose-Einstein
condensates have enabled experiments on a broad range of topics including
Feshbach resonances~\cite{ISI:000072462700052}, spinor
condensates~\cite{ISI:000221037500003}, and many-body effects in optical
lattices~\cite{BlochDalibardZwerger-2008}. Moreover, all-optical
schemes~\cite{AllOpticalBEC-Barrett-01} have eliminated the need for magnetic
trapping in evaporative cooling applications. Nevertheless, the use of magnetic
traps remains attractive due to their large volume, depth and passive
stability. Major recent developments include the advent of miniaturized
chip-based magnetic traps~\cite{Schmiedmayer-AtomChipReview-2002} and the
realization of transporter apparatus
designs~\cite{Greiner-01,Lewandowski-03,OptTweezersBEC-Gustavson-01,%
LargeBECMachines-Streed-06} in which the main chamber used for laser cooling is
spatially separated from the science cell in which experiments with the
condensate are performed.

Two generic types of transporter apparatus can be distinguished. In one type,
an evaporatively cooled cloud in or near the quantum degenerate regime is first
produced in the main chamber and subsequently delivered into the science cell
using an optical
tweezer~\cite{OptTweezersBEC-Gustavson-01,LargeBECMachines-Streed-06}. In the
second type, a laser-cooled, magnetically trapped cloud is first moved into the
science cell, before evaporation is performed. Here, the transport is achieved
by either using a chain of overlapping stationary coil pairs~\cite{Greiner-01},
or by mechanically displacing a single, movable pair of
coils~\cite{Lewandowski-03}. Transporter machines have become increasingly
popular, and both the optical and magnetic types are now in use by a number of
groups.

In comparing the two apparatus types, a certain complementarity can be noticed.
The optical tweezer design allows for nearly unobstructed optical access in the
science cell, but it requires the ability to perform evaporative cooling
already in the main chamber, thus imposing stringent conditions on the vacuum
and complicating apparatus design. In contrast, magnetic transporters only
require a simple vapor cell for laser cooling, but the presence of a permanent
magnetic coil assembly for evaporation restricts optical access in the science
cell.

In this paper, we present a novel implementation of a magnetic-coil transporter
apparatus, based on moving coils, that shares the advantages of the optical
tweezer scheme and also minimizes the overall design complexity: by adding a
homogeneous rotating bias field to that of the movable coil pair, a stable trap
of the TOP type~\cite{Petrich-95}, well-suited for condensate production, is
formed in the science cell. Since the same quadrupole coil pair is used for
magneto-optical trapping, transport, and evaporation, the TOP trap can thus be
operated as a ``retractable funnel'' to load an optical trap, resulting in
almost unobstructed optical access.

Hybrid traps, realized by the addition of either a
blue~\cite{Davis-95,NaikRaman-05} or red~{\cite{Lin-NewPortoMachine-2009}
detuned laser beam to the quadrupole potential, could present an alternative to
the solution presented here in the context of a moving-coil magnetic
transporter.  However, retaining the ability to produce condensates in an
alignment-free configuration seems advantageous for stable day-to-day
operation. Also, we are able to efficiently load condensates into optical traps
without incurring the need for high beam intensities.

This paper is organized as follows: Section II gives an overview of the
experimental setup, before the general features and technical implementation of
the TOP trap are discussed in section III. Section IV characterizes the
performance of the trap for condensate production, and section V illustrates
its use in conjunction with optical trapping applications.

\section{Overview of experimental setup}

\begin{figure}[th]
    \centering
    \includegraphics[width=7.3cm]{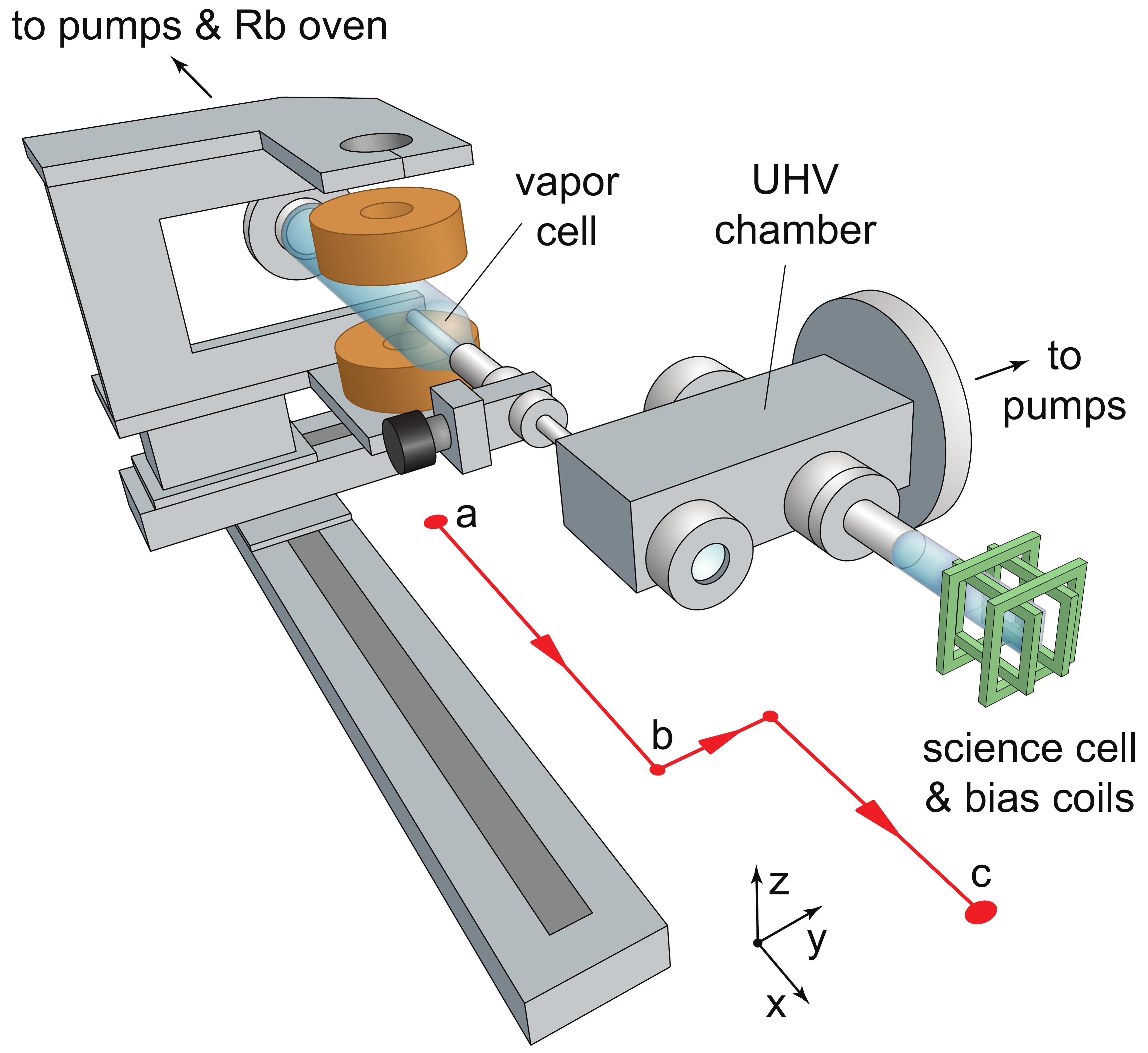}
    \caption{Transporter apparatus. The vapor cell is made from a 5.7~cm-diameter Pyrex tube
and is connected through valves with an ion pump and a rubidium reservoir (not
shown), maintaining a rubidium background pressure of several $10^{-9}$~torr.
A 28~cm-long tube of 1~cm inner diameter connects the vapor cell to the UHV
chamber and the attached science cell, which are pumped down to less than $10^{-11}$~torr by
an ion pump in combination with a titanium sublimation pump (not shown). The
quadrupole coils (shown in MOT position `a') are mounted on an aluminum
holder (sliced to reduce eddy currents) that sits on two orthogonally
stacked translation stages (Parker Daedal 404XR). The kink in the translation
path protects the science cell from fast
ballistic atoms escaping from the vapor cell, and it enhances optical access to
the condensate along the $x$-axis. A small gate valve located midway in the
pumping tube allows to completely disconnect the two vacuum regions for
servicing.}
    \label{fig:setup}
\end{figure}

Our moving-coil transporter apparatus is illustrated in figure~\ref{fig:setup}.
A cylindrical glass vapor cell is connected to a UHV chamber with an attached
small quartz glass science cell (cf.\ also figure~\ref{fig:McTOP}) through a
differential pumping tube. In the vapor cell, we collect up to $1.5\times
10^{10}$~atoms in a standard six-beam $^{87}$Rb MOT, making use of light
induced atom desorption
(LIAD)~\cite{LIAD-Anderson-01,LIAD-Klempt-06,Nakagawa-Transporter-2005} to
temporarily enhance the loading rate. After an 8~ms molasses phase that lowers
the temperature to about 25~$\mu$K, the atoms are optically pumped to the
$|F=1, m_F=-1\rangle$ hyperfine ground state, and subsequently caught in a
magnetic quadrupole trap using the same coil pair as for the MOT at an axial
field gradient of 100~G/cm, yielding $2.2\times10^{9}$ trapped atoms at
150~$\mu$K.

The quadrupole coils are mounted on a mechanical translation stage assembly,
which is used to move the magnetically trapped cloud into the science cell.
This is done as quickly as possible ($a=4.5$~m/s$^2$,
$v_{\mathrm{max}}=0.94$~m/s) in order to minimize losses due to collisions with
background gas atoms in the vapor cell. The quadrupole trap is simultaneously
compressed to its final axial field gradient of $B'_z = 350$~G/cm within 150~ms
before the cloud reaches the pumping tube. This adiabatically heats up the
cloud to 450~$\mu$K. We have confirmed that non-adiabatic heating during the
motion amounts to less than 10~$\mu$K.

Once the cloud reaches the UHV chamber~`b', where the measured lifetime exceeds
150~s, the transporter slows down and proceeds along a kinked path
($a=0.5$~m/s$^2$, $v_{\mathrm{max}}=0.18$~m/s). After a total travel time of
3~s and a covered distance of 66~cm, about $1.6\times 10^9$~atoms, or $75\%$ of
the atoms initially caught, arrive in the science cell~`c', where they are
evaporatively cooled as discussed in section~\ref{sect:McTOPBec}. We attribute
the loss of atoms during the motion to background gas collisions while moving
out of the vapor cell, and to a possible shaving off of hot atoms on the walls
of the differential pumping tube.

\section{Moving-coil TOP trap}

Instead of producing the condensate in a separate Ioffe-Pritchard trap, as
generally found in transporter apparatus based on~\cite{Lewandowski-03}, we use
the movable quadrupole coils as an integral part of the final magnetic trap.
This could be done, in principle, by realizing either a TOP
trap~\cite{Petrich-95} or a QUIC trap~\cite{QUICtrap-Esslinger-98}. For the
latter, however, the trap bottom depends on the delicate cancelation of the
much larger fields of the quadrupole coils and the Ioffe coil, making the trap
very sensitive to fluctuations of their relative positions.

For a TOP trap, the trap bottom is solely determined by the magnitude $B_0$ of
the rotating bias field. It is therefore inherently insensitive to the relative
positioning of the coils provided that the bias field is sufficiently
homogeneous. With the bias field rotating in the $xy$-plane, the time-averaged
magnetic potential at the center of the trap is given by
\begin{equation}
    V(\rho, z) = \mu B_0 + \case{1}{2} m\, \omega_{\!\perp}^2 \, \rho^2 + \case{1}{2} m\, \omega_z^2\, z^2
    \quad (\rho, z \ll \rho_0).
\end{equation}
Here, $m$ denotes the atomic mass, $\mu$ the magnetic moment, and $\rho_0 = B_0
/ B'_{\!\perp}$ is the radius of the ``circle-of-death'' on which the
field-zero is moving~\cite{Petrich-95}, where $B'_{\!\perp} = B'_{z}/2$ is the
radial quadrupole field gradient. The radial and axial trap frequencies are
$\omega_{\!\perp} = B'_{\!\perp} ( \mu / 2 m B_0 )^{1/2}$ and $\omega_z =
\sqrt{8}\: \omega_{\!\perp}$, respectively. A further advantage of a TOP trap
is that magnetic field fluctuations on time scales much slower than the trap
frequencies do not affect the trap bottom, unlike for dc magnetic traps. Our
``moving-coil TOP trap'' (McTOP) is formed by combining the movable quadrupole
coils with stationary bias-field coils at the science cell, as illustrated in
figure~\ref{fig:McTOP}. We note that a similar strategy has been used
in~\cite{TOPwaveguide-Reeves-05} in conjunction with a magnetic waveguide.

\begin{figure}[tbh]
    \centering
    \includegraphics[width=7.3cm]{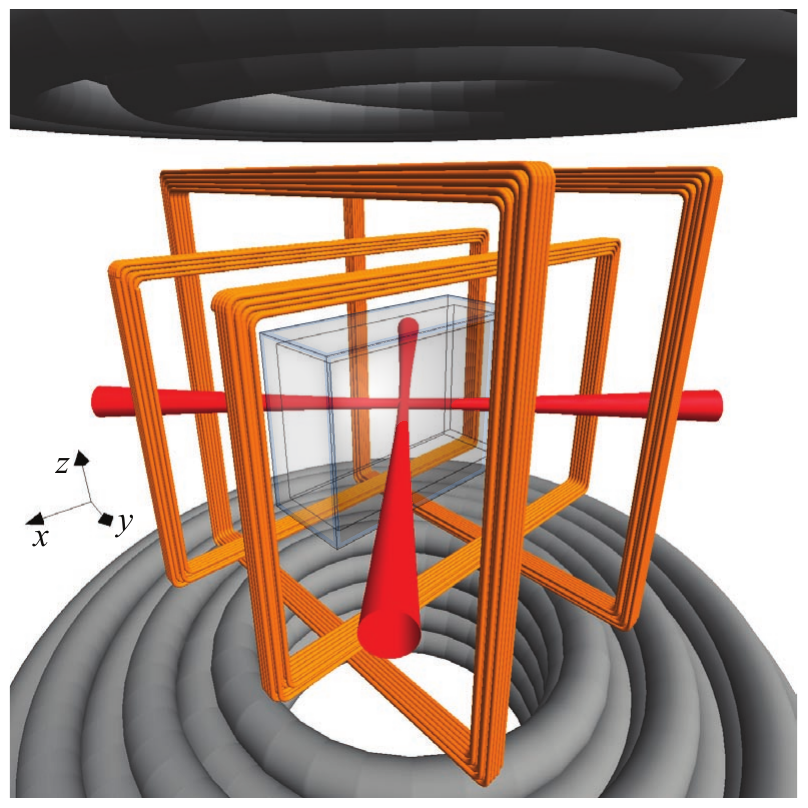}
    \caption{Moving-coil TOP trap configuration (McTOP) around the science cell. The
rectangular bias-field coils have dimensions of 56~mm$\times$37~mm (inner
coils) and 54~mm$\times$54~mm (outer coils), and a center-to-center spacing of
25.5~mm and 29.5~mm, respectively. They are tightly sandwiched between the
movable quadrupole coils used for transport (6.0~cm vertical clearance), which
each have an inner (outer) diameter of 4.4~cm (11.5~cm) and a height of 4.0~cm.
The science cell is a small quartz glass cell with inner dimensions of 10$\times$
20$\times$45~mm$^3$ and 1.25~mm wall thickness, which is fused to a
glass-to-metal adapter (not shown). Also sketched are the laser beams
forming the crossed optical dipole trap as described in section V.}
    \label{fig:McTOP}
\end{figure}

Each of the water-cooled quadrupole coils consists of 33~turns of
$1/4$~inch-diameter coated hollow copper tubing. At a current of 425~A, the
quadrupole coils ($L=90~\mu$H) produce an axial field gradient of $B'_z =
350$~G/cm , which can be switched off completely within less than 1~ms using an
IGBT. The bias-field coils are designed to provide a maximally homogeneous bias
field, while minimally obstructing the optical access to the trap center for
the given science cell geometry, as shown in figure~\ref{fig:McTOP}. The
outer~$B_x$ (inner $B_y$) coils each consist of 25 (20) turns of 24~AWG
(0.5~mm-diameter) magnet wire and are wound onto a stiff fiberglass holder
structure. Each coil pair produces a field of 7.5~G/A at the center of the trap
with a simulated field inhomogeneity of less than $\pm 3\times 10^{-4}$ within
a distance of 1~mm from the center. The air-cooled coils can thermally
withstand ac currents of 8~A amplitude for the duration of the evaporation,
corresponding to a 60~G bias field and a circle-of-death radius $\rho_0=3$~mm
at the maximum quadrupole field.

To drive the bias-field coils, we use an 800~W audio power amplifier~(PA). For
each coil pair, the impedance at the 10~kHz driving frequency is minimized down
to the ohmic resistance by canceling the inductance with a matching capacitor.
The measured inductance of the outer (inner) coil pair is $190~\mu$H
($110~\mu$H). The PA can easily supply the ac currents for a 60~G bias field
directly into the resulting loads of $\sim 1~\Omega$, without the need for
step-up transformers. The amplitude of the ac~current through each coil pair is
actively stabilized to within {$\sim10^{-4}$} using the circuit outlined in
figure~\ref{fig:TOPelectronics}. The regulation bandwidth of about 200~Hz is
more than sufficient for compensating thermal drifts in the coil
resistance~\footnote{An alternative, fast feedback scheme has been described
in~\cite{ACcurrentSource-Baranowski-06}, where the focus is on reducing current
noise to improve the coherence time for condensate interferometry in a special
TOP waveguide~\cite{TOPwaveguide-Reeves-05}.}. The current can be switched off
within less than 1~ms limited by the quality factor of the matched coil pair.

\begin{figure}[!b]
    \centering
    \includegraphics[width=8.6cm]{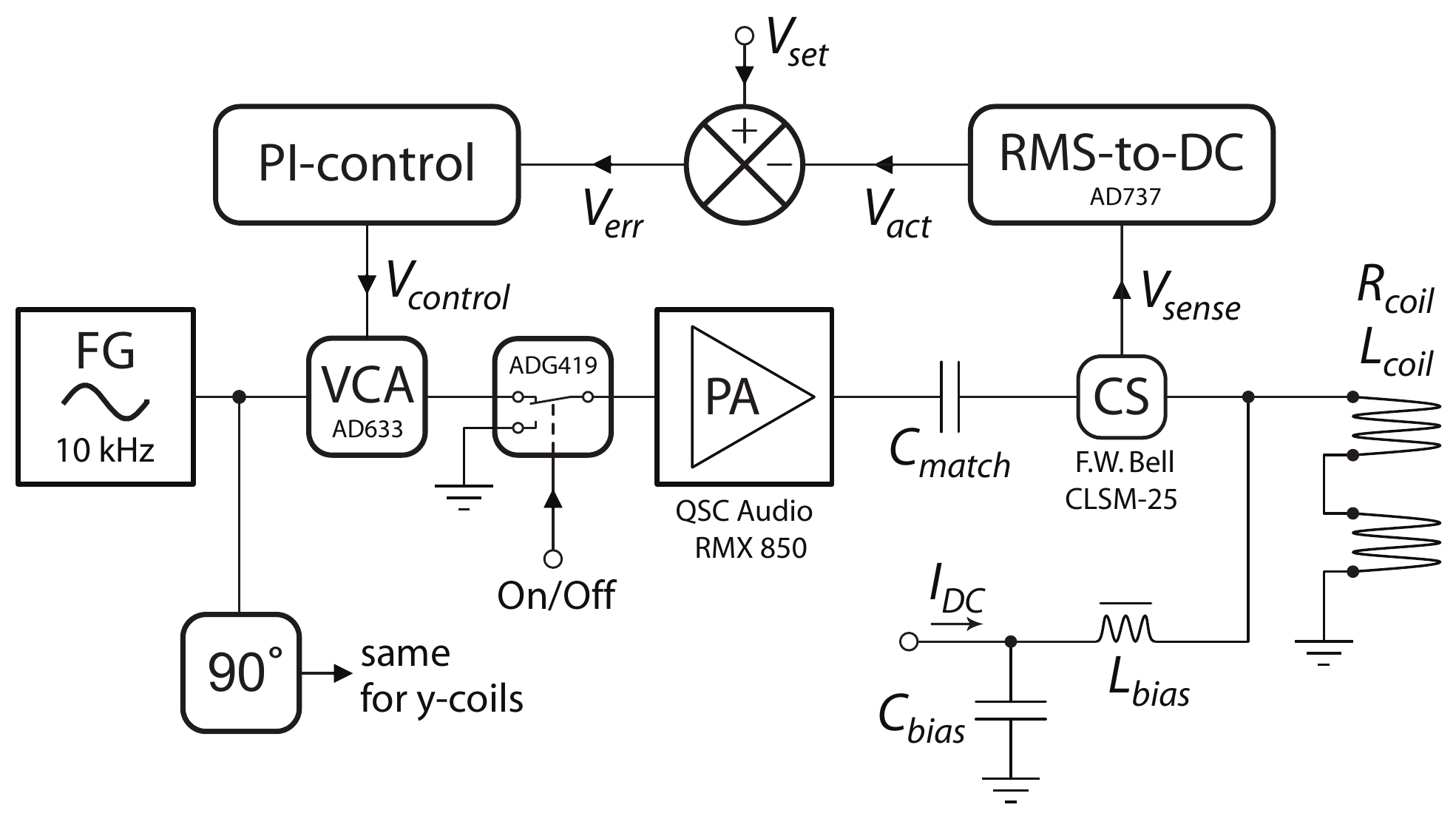}
    \caption{Closed-loop bias coil driving circuit (shown for one coil pair). An audio power
amplifier (PA) resonantly drives the coils at 10~kHz. The coil current is
sensed by a closed-loop Hall-effect current sensor~(CS) whose output is
converted into a dc~voltage~$V_\mathrm{act}$, corresponding to the amplitude of
the ac~current, by an rms-to-dc converter which determines the regulation bandwidth of
about 200~Hz. The error signal $V_\mathrm{err} = V_\mathrm{set}-V_\mathrm{act}$ is
fed into an op-amp integrator that acts as a
PI-controller and adjusts the gain of the voltage-controlled amplifier~(VCA) to
counteract any deviations from the desired amplitude~$V_\mathrm{set}$. The ac
current can be switched off rapidly with an analog switch. A bias-T
allows dc~currents to be run through the coils independent of the ac operation, e.g.\
for earth-field compensation.}
    \label{fig:TOPelectronics}
\end{figure}

\section{Condensate production in the M\lowercase{c}TOP trap}\label{sect:McTOPBec}

\begin{figure*}[t]
    \includegraphics[width=13cm]{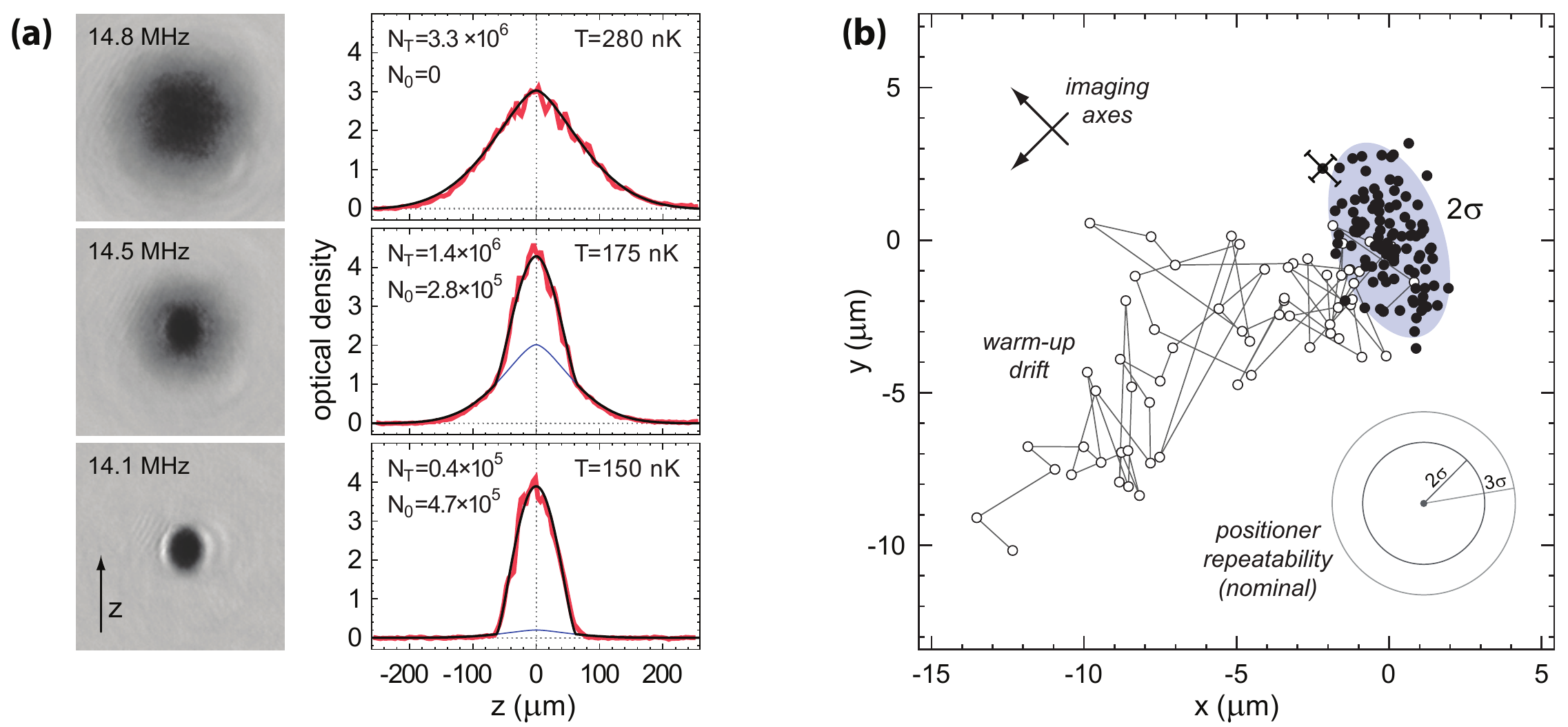}
    \caption{Condensate production in the McTOP trap: (a)~Phase transition as seen
in near-resonant absorption images after 16~ms time of flight along with
vertical cuts through the density profiles. The data are fitted with the sum of
two-dimensional Bose-enhanced Gaussian (thin line) and Thomas-Fermi
distributions~\protect\cite{Ketterle-VarennaBEC}. Condensation sets in at a
critical temperature of $(223\pm12)$~nK at which the cloud contains
$(2.7\pm0.4)\times10^6$ atoms. (b)~Reproducibility of the condensate position
in the horizontal plane. The filled circles show the in-situ positions of the
condensate (as determined from simple Gaussian fits) for the last 110 of a
series of 173 runs. The $2\sigma$ ellipsoid (shaded) with $95\%$ of the runs
has half-lengths of $3.3~\mu$m and $1.8~\mu$m, comparable to the specified
repeatability of the translation stages. The open circles represent a warm-up
drift during the first 63 runs (see text). The arrows indicate the two imaging
directions used, and the error bars indicate the maximum uncertainty in the
fits used to determine the condensate position. The radial Thomas-Fermi
diameter of the trapped condensate is $\sim40~\mu$m.}
    \label{fig:MTBEC}
\end{figure*}

After transport into the science cell, the atom cloud typically contains
$1.6\times10^{9}$ atoms in the $|1,-1\rangle$ state at a phase-space density of
$5\times10^{-7}$. Forced radio-frequency (rf) evaporative cooling is initially
performed in the stiff linear potential of the fully-compressed quadrupole trap
(350~G/cm axial gradient), where it is more efficient until Majorana losses
outweigh the advantage of a linear potential~\cite{KetterleDruten-96}. After a
14~s-long linear rf evaporation ramp down to a temperature of 75$~\mu$K and an
atom number of $7\times10^7$, the phase-space density has increased to $1\times
10^{-4}$ and the lifetime in the quadrupole trap due to Majorana losses has
decreased to 35~s. At this point, the trap is converted into a TOP trap by
switching on an 18~G rotating bias field, which preserves the atom number to
within $15\%$ and the phase-space density to within a factor of two. The
resulting harmonic trapping potential has measured trap frequencies of 70.4~Hz
in the axial and 25.0~Hz in the radial direction. The trap parameters are held
constant for the remaining 30~s of the evaporation sequence, during which
another piecewise-linear rf ramp takes the cloud to quantum degeneracy, cf.\
figure~\ref{fig:MTBEC}(a).

We have found the shot-to-shot position reproducibility of the condensate to be
consistent with the specified positioning uncertainty of the translation stages
($3\sigma=3\mu$m) once the system is warmed up, as shown in
figure~\ref{fig:MTBEC}(b). For this measurement, the condensate was imaged
simultaneously along two orthogonal axes in the horizontal $xy$-plane with
standard resonant absorption imaging on the repump transition, immediately
after the magnetic trap had turned off. We have not observed any systematic
shifts of the condensate position caused by the ``dual'' imaging itself  for
the beam intensities used.

During the first~63 of a total of 173~consecutive runs after a cold start, the
condensate position drifts by roughly~$15~\mu$m, as seen in
figure~\ref{fig:MTBEC}(b). In the $z$-direction we find a similar drift of
about 7~$\mu$m. This initial drift is directly correlated with the slow
temperature increase and subsequent stabilization at $\sim75^\circ$C of the
supply cables (4/0 AWG, i.e.\ 11.6~mm core diameter) of the quadrupole coils.
As the cables warm up, their thick rubber insulation becomes much less rigid,
thus changing the mechanical torque exerted on the coil holder. This potential
problem can most easily be avoided by pre-warming the cables at high quadrupole
coil currents.

We have found the condensate atom number to be stable to within $5$-$10\%$
depending on the performance of the MOT. For optimized conditions, condensates
containing up to $1\times10^6$ atoms have been observed. No correlations were
found between the atom number and the position jitter of the cloud in the
science cell.

\section{A retractable funnel for optical trapping}

\begin{figure}[t]
    \centering
    \includegraphics[width=7.3cm]{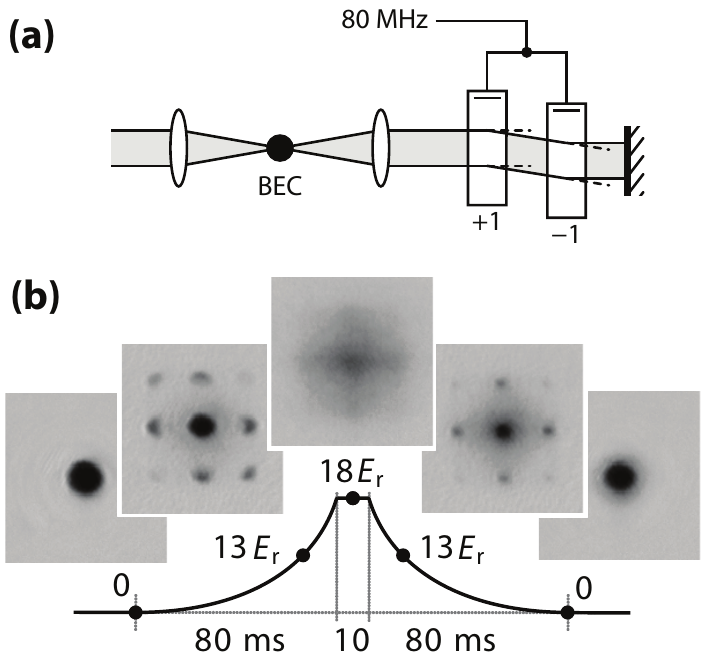}
    \caption{All-optical trapping and manipulation of a condensate after transfer from the
McTOP trap. (a)~Conversion of each of the two beams of the crossed dipole trap
(cf.\ figure~\ref{fig:McTOP}) into an optical lattice beam by partial
retro-reflection, using a double-pass AOM configuration~\protect\cite{Gemelke-2009}
with zero net frequency shift and rapidly adjustable
reflectivity of {$10^{-6}$-$10^{-1}$}. To realize a three-dimensional lattice, a
third beam pair (with full retro-reflection) is added along $z$.
(b)~Superfluid-to-Mott insulator transition in a three-dimensional optical lattice
as observed in absorption images after 18~ms time-of-flight. The lattice depth
follows exponential ramps of 80~ms duration, separated by a 10~ms hold time at
a depth of 18~recoil energies (solid line).}
    \label{fig:ODTOL}
\end{figure}

For condensate production in an optical trap, the evaporation in the McTOP trap
can also be used as an intermediate step after which the quadrupole coils are
moved out of the way. The stationary bias-field coils can then still be used to
control the spin quantization axis, for example.

The experimental procedure is as follows. After RF~evaporation in the magnetic
potential, the atoms are loaded adiabatically into a crossed-beam optical
dipole trap formed by two orthogonally intersecting laser beams. This is done
by smoothly ramping up the optical potential over 400~ms and then smoothly
ramping down the magnetic confinement over another 400~ms. After the transfer,
the quadrupole coils are moved back to the intermediate position~`b' indicated
in figure~\ref{fig:setup}. The Gaussian laser beams of the crossed optical
dipole trap (cf.\ figure~\ref{fig:McTOP}) have a $1/e^2$ radius of
$\sim135~\mu$m and a combined power of 3~W. They are derived from a
single-frequency 1064~nm ytterbium fiber laser (IPG YLR LP-SF series) with a
relative frequency offset of 20~MHz to average out interference effects. The
depth of the optical trap, including gravity, is $6~\mu$K in the horizontal and
$1~\mu$K in the vertical direction. This allows for efficient gravity-assisted
evaporation of atomic clouds. We typically load the optical trap with clouds at
$\sim250$~nK and then ramp down the trap depth to $5~\mu$K in the horizontal
and $200~$nK in the vertical direction where condensation sets in. At this
point, the trap is nearly isotropic with measured frequencies between 50 and
60~Hz. The loading procedure results in nearly pure condensates with atom
numbers that are within 90\% of those reached in the McTOP trap.

By performing the final evaporation in an optical trap, it is possible to
easily use the TOP trap for optical lattice experiments, thus avoiding the
usual drawback of TOP traps: atomic micromotion~\cite{Micromotion-Mueller-00}
can lead to strong heating due to an oscillatory motion of the atoms relative
to the lattice at the frequency of the rotating bias
field~\cite{Cristiani-LatticeinTOP-2002}. To demonstrate the suitability of the
approach presented in this paper, figure~\ref{fig:ODTOL} shows data for the
superfluid-to-Mott insulator transition in a
three-dimensional optical lattice~\cite{Jaksch-OptLatBHM-1998,%
Greiner-SF-MI-2002} obtained with our apparatus.

\section{Conclusion}

We have demonstrated a versatile and simple scheme for producing magnetically
and optically-trapped condensates in a moving-coil transporter apparatus. In
our scheme, the movable quadrupole coils are also used as an essential part of
the final magnetic trap. As a stand-alone device, this trap reliably produces
condensates with minimal technical complexity. In experiments with
optically-trapped Bose-Einstein condensates, the quadrupole coils can be
retracted before quantum degeneracy is reached, providing large optical access.
The apparatus is well-suited for experiments with optical lattices, as
demonstrated by observing the superfluid-to-Mott insulator transition.

\ack We thank R.~D. Schiller, D.~E. Sproles, B.~Bogucki, H.~Ruf, and A.~Hansen
for contributions in the early stages of the experiment, and R.~Reimann for
contributions to the optical lattice implementation. This work was funded by
the Research Foundation of SUNY, the Office of Naval Research (DURIP program),
and fellowships from the Fulbright program of the U.S. Department of State
(S.~A.) and from the GAANN program of the U.S. Department of Education (B.~G.).


\section*{References}




\end{document}